\documentclass{edbk} 
  \usepackage{edbkps}
\usepackage{epsfig}

\def\etal{{\it et al.}}

\def\eg{{\it e.g.}}
\def\lap{\hbox{${_{\displaystyle<}\atop^{\displaystyle\sim}}$}}
\def\gap{\hbox{${_{\displaystyle>}\atop^{\displaystyle\sim}}$}}

\let\footnote\savefootnote

\let\footnotetext\savefootnotetext 

\let\footnoterule\savefootnoterule 

\normallatexbib
\begin{document}



\articletitle[Probing the Neutron Star Interior]{Probing the Neutron
Star \\ Interior with Glitches}


\author{Bennett Link}
\affil{Montana State University\\
Los Alamos National Laboratory}
\email{blink@dante.physics.montana.edu}

\author{Richard I. Epstein}
\affil{Los Alamos National Laboratory}
\email{epstein@lanl.gov}

\author{James M. Lattimer}
\affil{State University of New York, Stony Brook}
\email{lattimer@astro.sunysb.edu}



\begin{abstract}

\noindent
With the aim of constraining the structural properties of neutron
stars and the equation of state of dense matter, we study sudden
spin-ups, {\em glitches}, occurring in the Vela pulsar and in six
other pulsars.  We present evidence that glitches represent a
self-regulating instability for which the star prepares over a waiting
time. The angular momentum requirements of glitches in Vela indicate
that $\ge 1.4$\% of the star's moment of inertia drives these
events. If glitches originate in the liquid of the inner crust, Vela's
`radiation radius' $R_\infty$ must exceed $\simeq 12$ km for a mass of
$1.4 M_\odot$. The isolated neutron star RX J18563-3754 is a promising
candidate for a definitive radius measurement, and offers to further 
our understanding of dense matter and the origin of glitches. 

\end{abstract}



\section{Introduction}

Many isolated pulsars suffer spin jumps, {\em glitches}, superimposed
upon otherwise gradual spin down under magnetic torque. For example,
in the glitch of the Crab pulsar shown in Fig. 1, the star spun up by
nearly a part in $10^7$ over several days \cite{crabglitch},
corresponding to a change in rotational energy of the crust of $\sim
10^{42}$ ergs. A particularly active glitching pulsar is the Vela
pulsar, which has produced more than a dozen glitches since its
discovery over 30 years ago. The fractional changes in rotation rate
are typically $\sim 10^{-6}$, occurring every three years on average
\cite{vela}.

Because glitching pulsars are isolated systems, glitches are thought
to arise from internal torques exerted by the rotating liquid interior
on the crust, whose spin rate we observe \cite{models}. As the star's
crust is spun down by the magnetic field frozen to it, the interior
liquid, which responds to the external torque indirectly through
friction with the solid crust, rotates more rapidly. For example, a
portion of the liquid could coast between glitches while the solid
crust spins down (Fig. 2). Glitches might arise as the consequence of
an instability that increases the frictional coupling between the
liquid and the solid, causing angular momentum flow to the crust.
\bigskip
 
\hspace*{-1.cm}
\epsfig{file=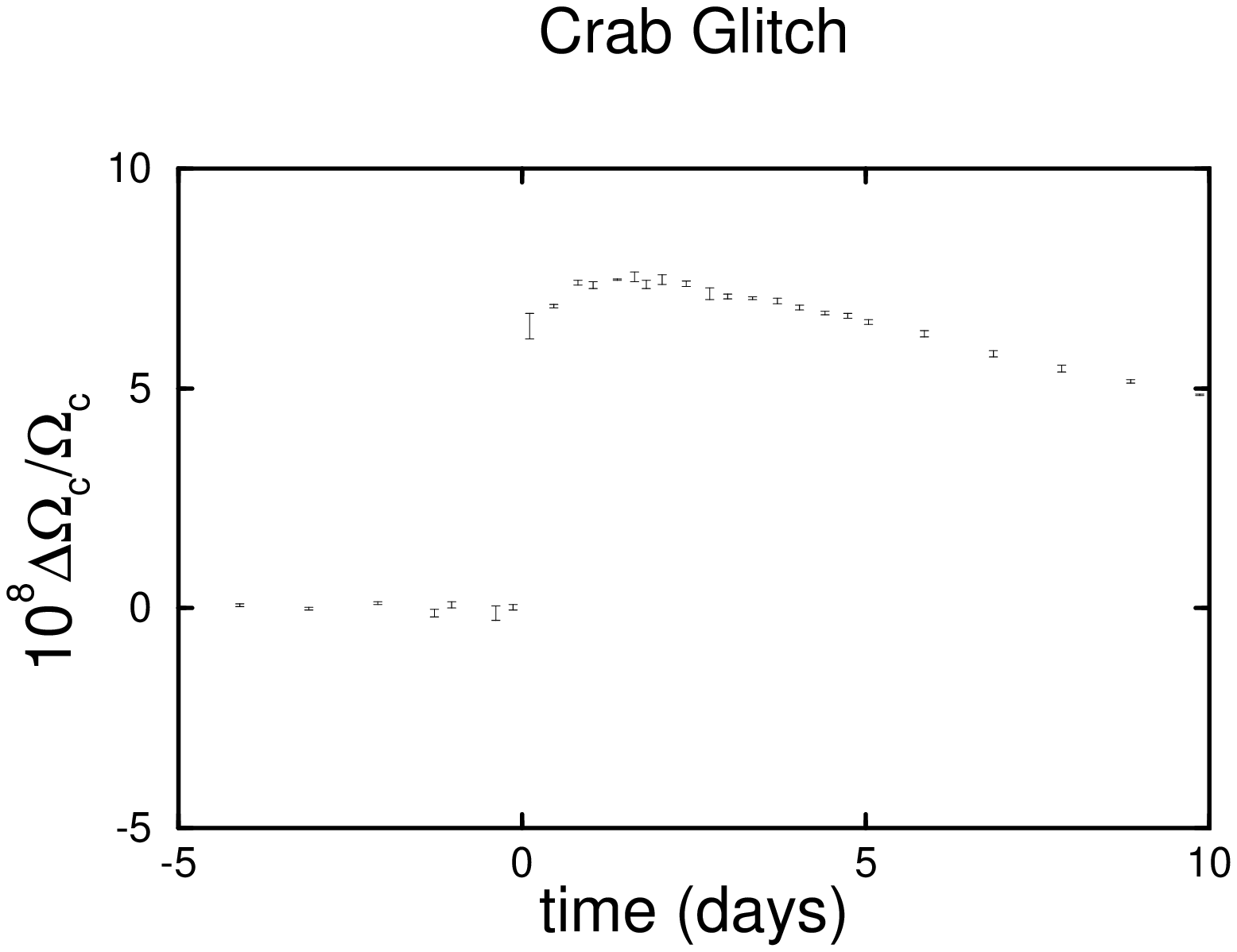,width=2.5in}
 
\vspace*{-1.8in}\hspace*{2.4in}
\epsfig{file=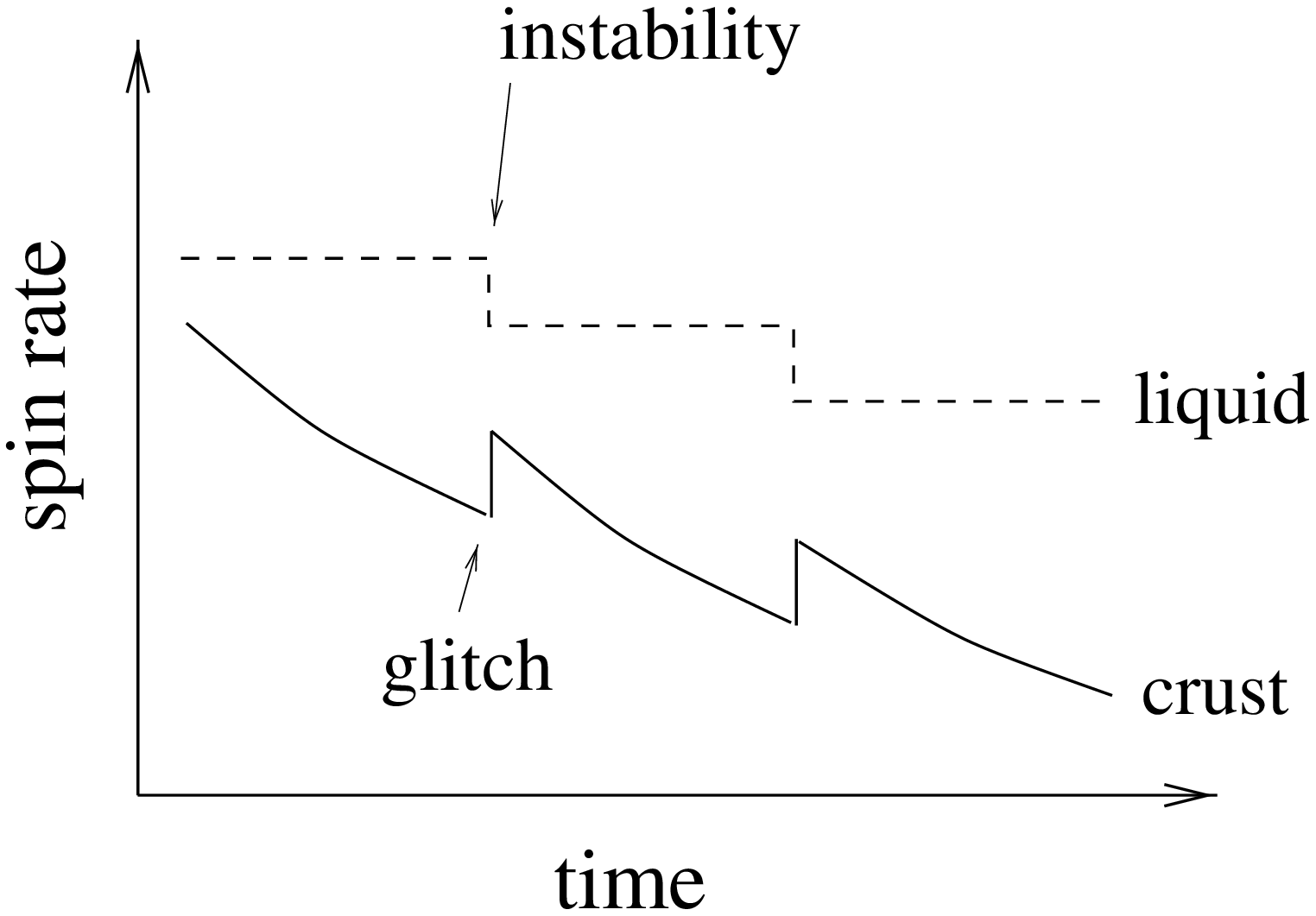,width=2.in}
  
\begin{figure}[ht]
\dblcaption{A glitch in the Crab pulsar. The spin rate with respect to
a pre-glitch spin-down model is shown. Data are from Ref. 1.}
{A glitch model. Differential rotation between the interior
liquid (dashed line) and the crust (solid line) develops as the crust
is spun down by its external torque. At the time of a glitch, the
liquid delivers angular momentum to the crust.}
\end{figure}

The long history of glitches in Vela makes it possible to deduce some
of the properties of the interior angular momentum reservoir
independent of the details of the instability that triggers these
events. Here we discuss the time distribution and average angular
momentum transfer rate of Vela's glitches and present evidence that
glitches in Vela represent a self-regulating instability for which the
star prepares over a waiting interval. We obtain a lower limit on the
fraction of the star's liquid interior responsible for glitches and
discuss how this result can be used to constrain the dense matter
equation of state and the structural properties of neutron stars. We
conclude with discussion of the nearby isolated neutron star RX
J185635-3754, a promising candidate for a robust radius measurement
that offers to constrain our understanding of dense matter and the
origin of glitches.

\section{Regularity of Angular Momentum Transfer}

A glitch of magnitude $\Delta\Omega_i$ requires angular momentum
\begin{equation}
\Delta J_i = I_c \Delta\Omega_i,
\end{equation}
where $I_c$ is the moment of inertia of the solid crust plus any
portions of the star tightly coupled to it.  Most of the core liquid
is expected to couple tightly to the star's solid component, so that
$I_c$ makes up at least 90\% of the star's total moment of
inertia \cite{core}. [Glitches are driven by the portion of the liquid
interior that is differentially rotating with respect to the crust].
The cumulative angular momentum imparted to the crust over time is
\begin{equation}
J(t) = I_c\bar{\Omega}\sum_{i}
 {\Delta\Omega_i\over\bar{\Omega}},
\end{equation}
where $\bar{\Omega}= 70.4$ rad s$^{-1}$ is the average spin rate of
the crust over the period of observations. Fig. 3 shows the cumulative
dimensionless angular momentum, ${ J}(t)/I_c\bar{\Omega}$, over $\sim
30$ years of glitch observations of the Vela pulsar, with a linear
least-squares fit.  The average rate of angular momentum transfer
associated with glitches is $I_c\bar{\Omega}A$, where $A$ is the
slope of the straight line in Fig. 3:
\begin{equation}
A =
(6.44\pm 0.19)\times 10^{-7}\,\,{\rm yr}^{-1}.
\label{jdot}
\end{equation}
This rate $A$ is often referred to as the {\it pulsar activity
parameter} (see, \eg, \cite{lyne}).

\begin{figure}[ht]
\epsfig{file=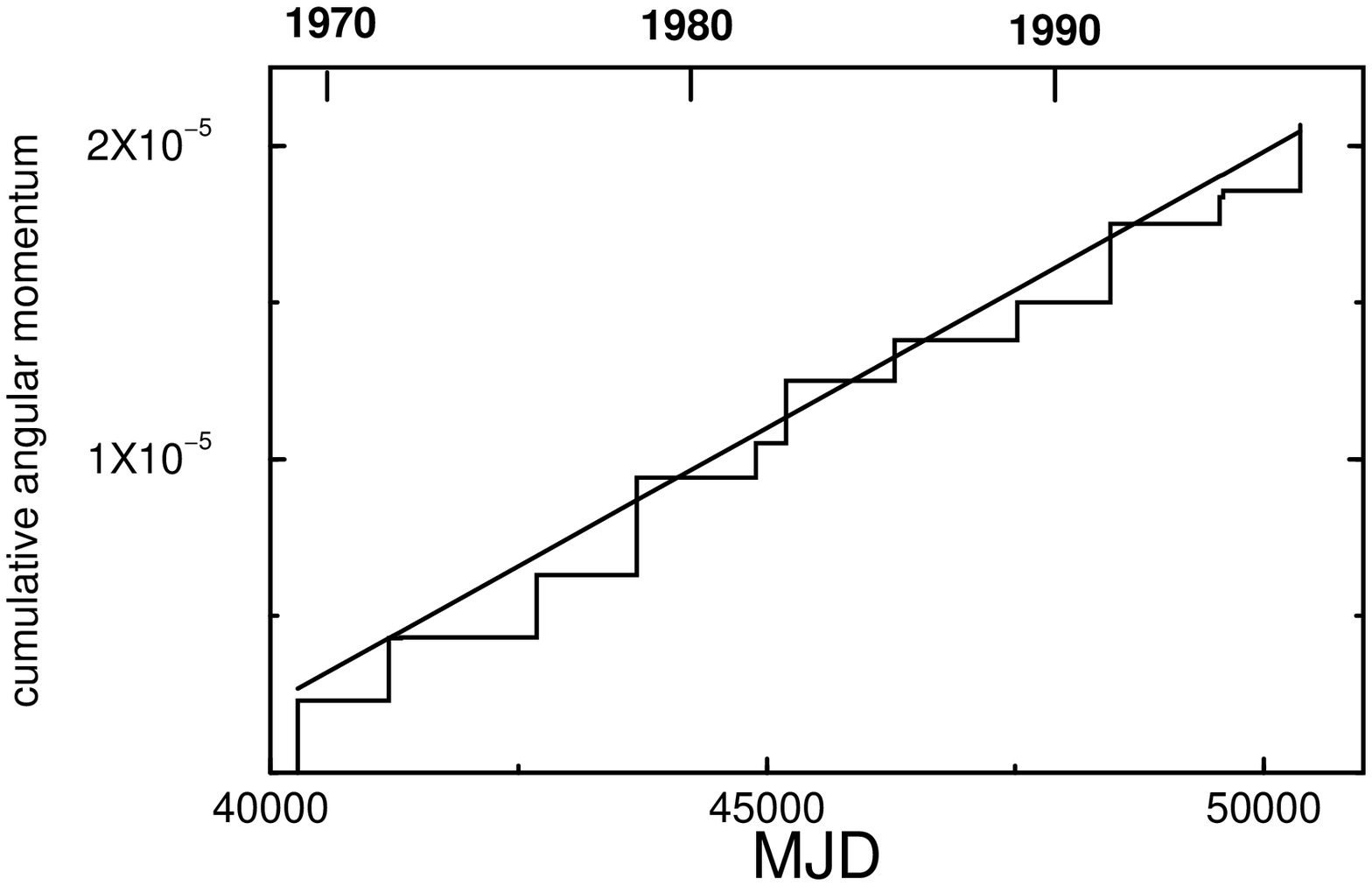,width=3.in}
\vspace*{-2.in}
\narrowcaption
{Cumulative dimensionless angular momentum,
$J/I_c\bar{\Omega}$, imparted to the Vela pulsar's crust as a function
of time. The straight line is a least-squares fit.}
\end{figure}
\vspace*{1.cm}

The angular momentum flow is remarkably regular; none of Vela's 13
glitches caused the cumulative angular momentum curve to deviate from
the linear fit by more than 12\%.  Moreover, glitches occur at fairly
regular time intervals; the standard deviation in observed glitch
intervals is $0.53\langle\Delta t\rangle$, where $\langle\Delta
t\rangle = 840$ d is the average glitch time interval. The probability
of 13 randomly-spaced (Poisson) events having less than the observed
standard deviation is only $\sim 1$\%. These data indicate that Vela's
glitches are not random, but represent a self-regulating process which
gives a relatively constant flow of angular momentum to the crust with
glitches occurring at fairly regular time intervals.

\section{The Glitch Reservoir's Moment of Inertia}

The frequent occurrence of large glitches requires that some fraction
of the interior liquid (the reservoir) spins at a higher rate than the
crust of the star.  The average rate of angular momentum transfer in
Vela's glitches (eq. \ref{jdot}) can be used to constrain the relative
moment of inertia of the reservoir. Between glitches, the reservoir
acquires excess angular momentum as the rest of the star slows under
the magnetic braking torque acting on the crust.  Excess angular
momentum accumulates at the maximum rate if the reservoir {\em coasts}
between glitches, without spinning down (Fig. 2). Hence, the rate at
which the reservoir accumulates angular momentum capable of driving
glitches is limited by
\begin{equation}
\dot{J}_{\rm res} \le I_{\rm res} |\dot{\Omega}|,
\end{equation}
where $\dot{\Omega}$ is the average spin-down rate of the
crust, and $I_{\rm res}$ is the moment of inertia of the angular momentum
reservoir (not necessarily one region of the star).
Equating $\dot{J}_{\rm res}$ to the average rate of
angular momentum transfer to the crust,
$I_c\bar{\Omega}A$, gives the constraint,
\begin{equation}
{I_{\rm res}\over I_c} \ge
{\bar{\Omega}\over\vert\dot{\Omega}\vert} A
\equiv G,
\label{constraint}
\end{equation}
where the {\em coupling parameter} $G$ is the minimum fraction of the
star's moment of inertia that stores angular momentum and imparts it
to the crust in glitches.  Using the observed value of Vela's activity
parameter $A$ and $\bar{\Omega}/\vert\dot{\Omega}\vert=22.6$ Kyr, we
obtain the constraint
\begin{equation} 
{I_{\rm res}\over I_c} \ge G_{\rm Vela}= 1.4\%.
\end{equation}
A similar analysis for six other pulsars yields the results shown in
Fig.  4. After Vela, the most significant limit is obtained from PSR
1737-30 which gives $I_{\rm res}/I_c\ge G_{\rm 1737}= 1$\%.

\hspace{-1.3cm}
\epsfig{file=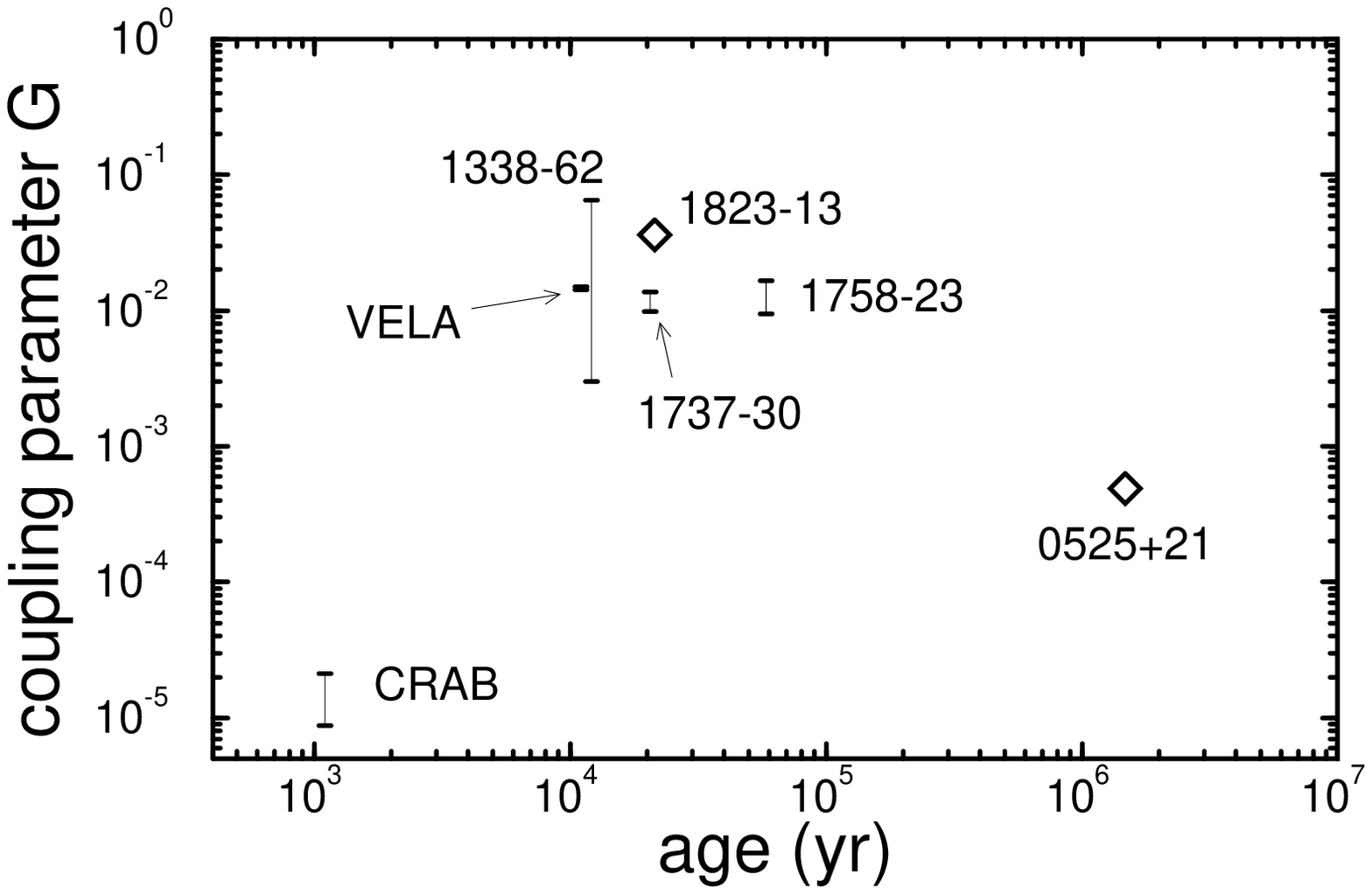,width=3.4in}
\begin{figure}[ht]
\vspace*{-2.7in}
\narrowcaption
{The coupling parameter G.  The strongest constraints are
obtained for Vela and PSR 1737-30, for which 13 and 9 glitches have
been observed, respectively.  Diamonds indicate objects with only two
observed glitches, for which error bars could not be obtained.
References:  0525+21 \cite{downs82}, Crab \cite{crab}, Vela \cite{vela}, 
1338-62 \cite{kaspi92}, 1737-30 \cite{ml,sl}, 1823-13 \cite{sl}.}
\end{figure}

The similarity of $G$ for the five objects of intermediate age
suggests a common physical origin for glitches in these stars.  The
Crab pulsar and PSR 0525+21, however, appear to be unusual.  It may be
that the Crab's interior cannot accumulate significant excess angular
momentum between glitches, perhaps as a consequence of rapid thermal
creep of superfluid vortices (see, e.g., \cite{creep}). The value of
$G$ for PSR 0525+21 is not well determined, since only two glitches
from this object have been measured.

\section{Implications for the Dense Matter Equation of State}

The constraint of $I_{\rm res}/I_c\ge 1.4$\% for Vela applies
regardless of where in the star glitches originate. A natural
candidate for the angular momentum reservoir is the superfluid that
coexists with the inner crust lattice \cite{models}, where superfluid
vortex lines could pin to the nuclei and sustain a velocity difference
between the superfluid and the crust. Within this interpretation, the
constraint on $I_{\rm res}/I_c$ can be used to constrain the
properties of matter at supranuclear density as we now demonstrate.

The fraction of the star's moment of inertia contained in the solid
crust (and the neutron liquid that coexists with it) is given
approximately by \cite{prl}:
\begin{equation}{\Delta I\over
I}\simeq {28\pi\over 3}{P_t R^4\over GM^2}
\left[1+{8P_t\over n_t
m_nc^2}{4.5+(\Lambda-1)^{-1}\over\Lambda-1}\right]^{-1}\,.
\label{integral}
\end{equation}
Here $n_t$ is the density at the core-crust boundary, $P_t$ is the
pressure there, $M$ and $R$ are the stellar mass and radius, 
$\Lambda\equiv(1-2GM/Rc^2)^{-1}$ is the gravitational redshift and
$m_n$ is the neutron mass.  $\Delta I/I$ is a function of $M$ and $R$
with an additional dependence upon the equation of state (EOS) arising
through the values of $P_t$ and $n_t$.  $P_t$ is the main EOS
parameter as $n_t$ enters chiefly via a correction term.  In general,
$P_t$ varies over the range $0.25<P_t<0.65$ MeV fm$^{-3}$ for
realistic equations of state \cite{lp}. Larger values of $P_t$ give
larger values for $\Delta I/I$.
\begin{figure}[ht]
\hspace*{-.1cm}
\epsfig{file=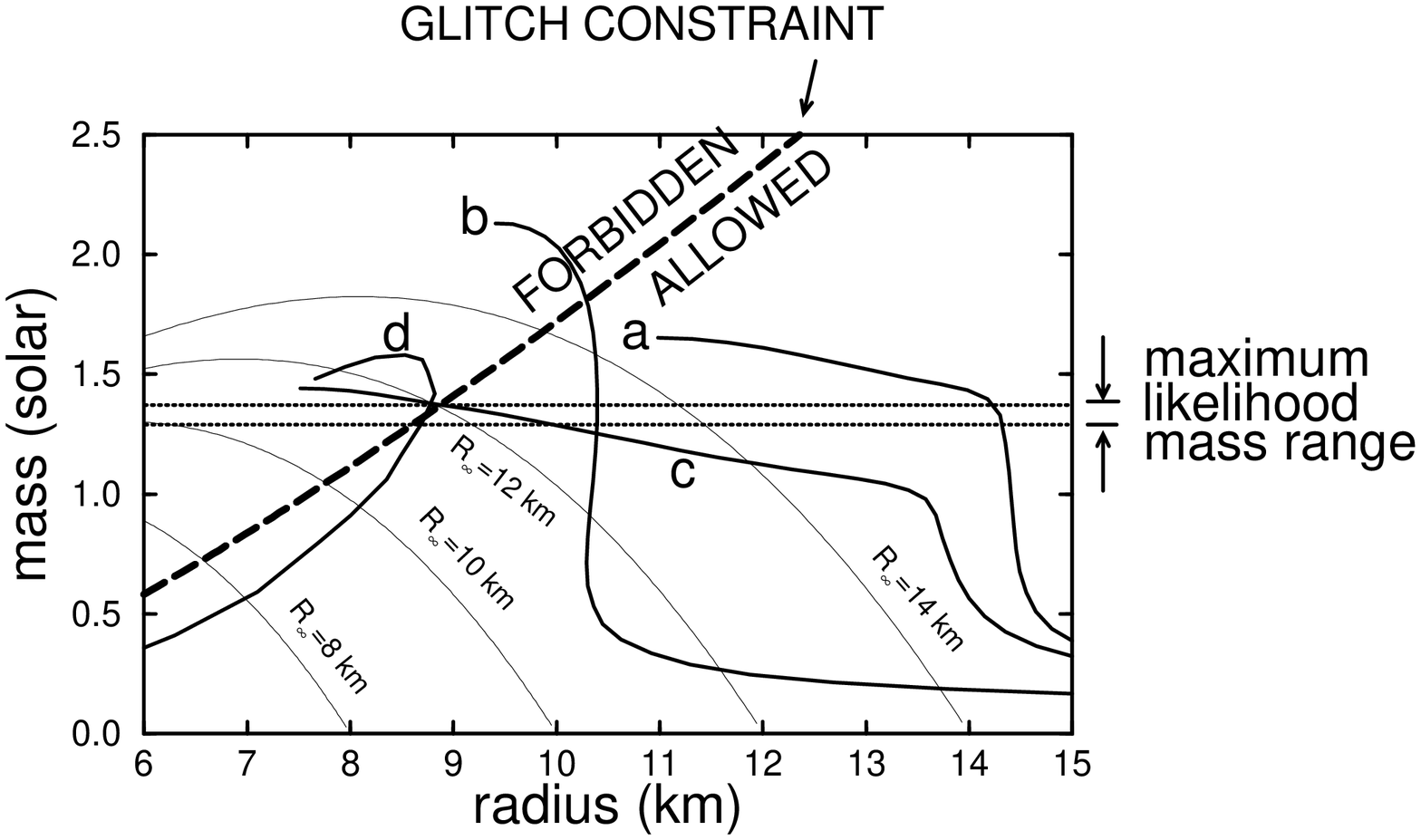,width=4.5in}
\vspace*{-1.5cm}\caption{Limits on the Vela pulsar's radius. The heavy
dashed curve delimits allowed masses and radii that are compatible
with the glitch constraint $\Delta I/(I-\Delta I)\ge 1.4$\% for
$P_t=0.65$ MeV fm$^{-3}$. The horizontal dashed lines indicate the
mass limits for the survey of 26 radio pulsars of Ref. 16. For
comparison, four mass-radius relations are shown for the equations of
state of: Glendenning \& Schaffner-Bielich (kaons, curve a) \cite{gs},
Wiringa, Fiks \& Fabrocini (curve b) \cite{wff}, Glendenning \&
Schaffner-Bielich (kaons, curve c) \cite{gs} and a representative
strange quark star (curve d).  Thin curves are contours of constant
radiation radius $R_\infty$. }
\end{figure}

Combining eq. [\ref{integral}] with a lower limit on $\Delta I$ and an
upper limit on $P_t$ gives a lower limit on the neutron star radius
for a given mass. In order to relate our observational bound on
$I_{\rm res}/I_c$ to $\Delta I$, we assume that the angular momentum
reservoir is confined to the neutron superfluid that coexists with the
nuclei of the inner crust. In this case, $I_{\rm res}\le\Delta I$ and
$I_c\ge I-\Delta I$. Our observational limit on $I_{\rm res}$ then
gives $\Delta I/( I-\Delta I)\ge \Delta I/I_c \ge I_{\rm res}/I_c \ge
0.014$. To obtain a conservative lower limit on the neutron star radius, we
take $P_t=0.65$ MeV fm$^{-3}$ and $n_t=0.075$
fm$^{-3}$. Eq. [\ref{integral}], gives the heavy dashed curve in
Fig. 5.  This curve is given approximately by
\begin{equation}
R=3.6+3.9 M/M_\odot\,.
\label{rlimit}
\end{equation}
Stellar models that are compatible with the lower bound on $I_{\rm
res}$ must fall below this line. For models above this line, the inner
crust liquid constitutes less than 1.4\% of the star's moment of
inertia, incompatible with the observed angular momentum requirments
of the Vela pulsar. 

\section{Discussion}

Vela's mass and radius are unknown, so we cannot say where it falls
in Fig. 5. However, mass measurements of radio pulsars in binary
systems and of neutron star companions of radio pulsars give neutron
star masses that are consistent with a remarkably narrow distribution,
$M=1.35\pm 0.04 M_\odot$ \cite{masses}, indicated by the pair of
horizontal dotted lines in Fig. 5. If Vela's mass is not unusual,
eq. [\ref{rlimit}] constrains $R\gap 8.9$ km, under the assumption
that glitches arise in the inner crust superfluid. However, the
quantity constrained by observations of the stellar luminosity and
spectrum is not $R$ but the larger `radiation radius'
$R_\infty\equiv\Lambda^{1/2}R =(1-2GM/Rc^2)^{-1/2}R$. If
$M=1.35M_\odot$ for Vela, the above constraint gives $R_\infty\gap 12$
km if glitches arise in the inner crust.  For comparison, we show in
Fig. 5 the mass-radius curves for several representative equations of
state (heavy solid lines). Measurement of $R_\infty\lap 13$ km would
be inconsistent with most equations of state if $M\simeq 1.35
M_\odot$. Stronger constraints could be obtained if improved
calculations of nuclear matter properties give $P_t$ significantly
less than 0.65 MeV fm$^{-3}$.\footnote{If $P_t$ is smaller than 0.65
MeV fm$^{-3}$, the crust's moment of inertia would be smaller and the
radius constraint more restrictive. For example, $P_t=0.25$ MeV
fm$^{-3}$ moves the constraining contour to approximately
$R=4.7+4.1M/M_\odot$.} For example, for $M\simeq 1.35 M_\odot$,
$R_\infty\gap 13$ km would be required if $P_t=0.25$ MeV fm$^{-3}$. A
measurement of $R_\infty\lap 11$ km would rule out most equations of
state regardless of mass or the angular momentum requirements of
glitches, and could indicate that neutron stars are not made of
neutrons at all, but of strange quark matter. Explaining glitches in
this case would be problematic, as strange stars have very thin crusts
\cite{strangestar}.

A black body fit to the unpulsed component of Vela's thermal emission
gives $R_\infty=3-4$ km \cite{ofz}. This result is difficult to
interpret without knowledge of the star's atmospheric composition and
magnetic field strength; atmospheric effects could increase this
estimate by a factor of up $\sim 6$, but could also decrease it by a
factor $\sim 2$ \cite{rrm}. Nevertheless, it would be interesting to
check the extent to which our constraints on Vela are obeyed by 
other neutron stars. A promising candidate for a decisive measurement
of a neutron star's radiation radius is RX J185635-3754, an isolated,
non-pulsing neutron star \cite{walter}. A black body fit to the X-ray
spectrum gives $R_\infty =7.3 (D/120\ {\rm pc})$ km where $D$ is the
distance (known to be less than 120 pc).  Taken at face value, this
result would not only be inconsistent with the radius requirements of
glitches in Vela - it would rule out all equations of state that do
not involve strange matter. [It is possible that glitching pulsars are
normal neutron stars, while RX J185635-3754 is `strange'].  However,
either a non-uniform surface temperature or radiative transfer effects
in the stellar atmosphere could raise this estimate significantly
\cite{alpw}. Recent HST observations of this source by F. Walter
should give the proper motion and parallax, and hence, the
distance. Future CHANDRA observations should yield more detailed
spectral data and could establish the composition of the atmosphere if
absorption lines are identified. If lines are present, atmospheric
fits will give $R_\infty$, $R$ and $M$, thus restricting RX
J185635-3754 to a region of Fig. 5. These data will undoubtedly
further our understanding of matter at supranuclear density, and could
establish whether neutron stars have properties consistent with an
inner crust explanation of glitches.



\begin{acknowledgments}

We thank A. G. Lyne for providing us with glitch data for the Crab
pulsar and P. M. McCulloch for Vela data.  This work was performed
under the auspices of the U.S.  Department of Energy, and was
supported in part by NASA EPSCoR Grant \#291748, NASA ATP Grants \#
NAG 53688 and \# NAG 52863, by the USDOE grant DOE/DE-FG02-87ER-40317,
and by IGPP at LANL.

\end{acknowledgments}



%

\bibliographystyle{apalike}


\end{document}